\input harvmac

\def\half{{1\over 2}}

\def\sh{\hat{\sigma}}



\Title{}{\vbox{\centerline{Macroscopic Black Holes, Microscopic Black Holes}
\bigskip
\centerline{and Noncommutative Membrane}}}

\centerline{Miao Li }
\centerline{\it Institute of Theoretical Physics}
\centerline{\it Academia Sinica, P.O. Box 2735}
\centerline{\it Beijing 100080, P. R. China}
\medskip
\centerline{\tt mli@itp.ac.cn}

\bigskip

We study the stretched membrane of a black hole as consisting of a perfect
fluid. We find that the pressure of this fluid is negative and the specific
heat is negative too. A surprising result is that if we are to assume the fluid
be composed of some quanta, then the dispersion relation of the fundamental
quantum is $E=m^2/k$, with $m$ at the scale of the Planck mass. There are
two possible interpretation of this dispersion relation, one is the noncommutative
spacetime on the stretched membrane, another is that the fundamental quantum is
microscopic black holes.

\Date{November, 2003}

\nref\sv{A. Strominger and C. Vafa, ``Microscopic Origin of the
Bekenstein-Hawking Entropy," hep-th/9601029, Phys.Lett. B379
(1996) 99.} \nref\thorne{K. S. Thorne, ``Black holes in the
membrane paradigm," Yale Univ. Press, 1986.}
\nref\qmem{ L. Susskind, L. Thorlacius, J. Uglum, hep-th/9306069,
Phys.Rev. D48 (1993) 3743;
L. Susskind, hep-th/9309145; G. 't Hooft,
Nucl.Phys. B256 (1985) 727; A. Sen, hep-th/9504147, Mod.Phys.Lett. A10 (1995)
2081.}
\nref\ty{T. Yoneya, in ``Wandering in Fields," eds. K. Kawarabayashi, A. Ukawa
(World Scientific, 1987); M. Li and T. Yoneya, Phys. Rev. Lett.
78, (1997) 1219, hep-th/9611072; T. Yoneya, Prog. Theor. Phys. 103
(2000) 1081, hep-th/0004074.} \nref\sh{S. Hawking, ``Particle
creation by black holes," Comm. Math. Phys. 43 (1975) 199.}
\nref\ms{S. R. Das and S. D. Mathur, ``Comparing decay rates for
black holes and D-branes," hep-th/9606185, Nucl.Phys. B478 (1996)
561. J. Maldacena and A. Strominger, ``Black Hole Greybody Factors
and D-Brane Spectroscopy," . hep-th/9609026, Phys.Rev. D55 (1997)
861.}
\nref\gl{Y. H. Gao and M. Li, hep-th/9810053, Nucl.Phys. B551 (1999) 229.}

Despite much progress made in the last decade in understanding the quantum properties
of black holes, in particular in string theory \sv, it is fair to say that we know
very little about what happens to the standard Schwarzschild black holes.
The difficulty in string theory to resolve the quantum black hole puzzle lies
in the fact that this is an intrinsic nonperturbative problem.

In this paper we take the stand-point of a phenomenological approach. We assume
the following: that the Bekenstein-Hawking entropy is correct, that there is
a Hawking radiation, and that there is a stretched horizon a proper distance away
from the real horizon \thorne. With this minimal input, we shall show that
the stretched horizon consists of a perfect fluid with a negative pressure, and
that the fundamental degree of the freedom in this fluid enjoys a strange dispersion
relation $E=m^2/k$, with $m\sim m_P$, $m_P$ is the Planck mass scale.
This relation suggests a noncommutative stretched horizon, with time noncommuting
with coordinates on the horizon. This relation has a second interpretation, if we
replace $k$ by $1/L$, where $L$ is the size a microscopic black hole, then the
dispersion relation is simply the relation between the mass of the microscopic
black hole and its size.

Let us start with a 4 dimensional black hole with the Schwarzschild metric
\eqn\fssm{ds^2=-(1-{r_0\over r})dt^2+(1-{r_0\over r})^{-1}dr^2+r^2d\Omega^2_2.}
The relation between the horizon radius and the mass of the black hole is
$2M=l_p^{-2}r_0$, where $l_p^2=G$ is the Planck length squared. The Hawking
temperature and the Bekenstein-Hawking entropy are
\eqn\hbh{T_H={1\over 4\pi r_0},\quad S={A\over 4\l_p^2},}
where $A$ is the area of the horizon.

Many people have studied the stretched horizon, a sphere a proper
distance away from the real horizon \thorne, the quantum properties of
the stretched horizon were studied in \qmem. Let the location of
the stretched horizon be $r_s>r_0$, very close to $r_0$. The
proper distance from the real horizon is
\eqn\pdis{l=\int_{r_0}^{r_s}g^{-1/2}_{00}(r)dr=2r_0g^{1/2}_{00}(r_s),}
namely, the ratio of $r_0$ and the proper distance is simply the
blue-shift factor: $2r_0/l=g_{00}^{-1/2}(r_s)$. Usually, based on
the black hole physics, people assume that the proper distance $l$
is close to the Planck scale. The exact formula for $l$ is given
by
\eqn\expdis{l=r_0(x+\half\sinh(2x)),\quad \sinh
x=\sqrt{{r_s\over r_0}}g_{00}^{1/2}(r_s).}
For a macroscopic black hole, $l/r_0$ is very small, since $l$ is around
the Planck scale, so $x$ is equal to $l/(2r_0)$ up to correction
$O((l/r_0)^3)$. Next, the second formula in \expdis\ tells us that
$x$ is equal to $g_{00}^{1/2}$ up to a correction $g_{00}x$.
Put these together, we find that the formula in \pdis\ is valid
up to a correction $(l/r_0)^3$ when rewritten as $g_{00}^{1/2}(r_s)
=l/(2r_0)$.

The stretched horizon has a temperature $T$ blue-shifted from the Hawking
temperature. Since $Tg^{1/2}_{00}(r_s)=T_H$, we have
\eqn\ht{T={1\over 2\pi l}.}
This relation is quite universal, valid for many types of black holes as well
as the de Sitter space. If we attribute the black hole mass completely to
the stretched horizon, since the mass is also blue-shifted, the mass
density on the stretched horizon is $Mg^{-1/2}_{00}/A$, namely
\eqn\mmd{\rho={1\over 4\pi l l_P^2}.}
This is a two dimensional density, so its dimension is $m^3$.
The density of entropy is simply $S/A$,
\eqn\ed{\sigma={S\over A}={1\over 4l_P^2}.}
Since on the membrane, the temperature is given by \ht, the relation
between the energy density and the entropy density is
\eqn\msd{\rho=2T\sigma.}
Or, $\rho/\sigma=2T$. Since $\sigma$ is the effective number of degrees
of freedom, this relation states that the energy per degree of freedom
is $2T$.

We already obtained a strange result. For a nonrelativistic noninteracting
gas, the energy per degree is $\half T$, this is also true for a
one-dimensional relativistic gas. In 2D which happens to be the case of
the stretched membrane, the energy per degree is $(2/3)T$. For a $d$
dimensional relativistic gas, we have $\rho/\sigma =(d/(d+1))T$. Thus,
we can never hope to get the result $\rho/\sigma=2T$ from any known
gas.

Assume that on the membrane there is a perfect fluid. This fluid, for
a sufficiently macroscopic black hole, must be universal, that is, its
properties do not depend on the mass of the black hole at all.
We believe that the stretched membrane has a fixed temperature and other
thermodynamic quantities, nevertheless we shall for now imagine that its
temperature, energy density and entropy density can be changed, in order
to use thermodynamics to gain some insight into this fluid.
We now apply the first law of thermodynamics to the membrane:
\eqn\flawt{d(\rho A)=Td(\sigma A)-pdA.}
If $\rho$ and $\sigma$ are functions of the temperature only, or physical
quantities independent of $A$, we obtain from \flawt\ relations ,
\eqn\twor{\rho=T\sigma -p,\quad d\rho = Td\sigma.}
From the first relation and \msd\ which we assume to be valid for any $T$
(certainly true for many known systems), we obtain
\eqn\pres{p=-T\sigma =-\half \rho.}
Thus, the pressure is indeed negative. The above relation is quite similar
to quintessence, but the latter is not a thermodynamic system at all.
The negative pressure may be responsible for the stability of the membrane
under self-gravity.

Relation \msd\ together with the second relation results in
\eqn\rst{\rho=2cT^{-1},\quad \sigma=cT^{-2}.}
These are certainly unusual relations. For $T\sim l^{-1}\sim l_P^{-1}$,
we find $c\sim l_P^{-4}=m_P^4$, for now, it is just a undetermined constant.
The specific heat is negative as implied by \rst
\eqn\sph{{d\rho\over dT}==-2cT^{-2}=-2\sigma,}
the same as for the black hole. Indeed, for the black hole itself
\eqn\sphh{{dM\over dT_H}=-2S.}
This similarity supports our scheme in which the membrane fluid is allowed
to change its temperature.

Next, we try to use statistical mechanics to gain some insight into the
microscopic nature of the membrane fluid. All thermodynamic quantities
can be obtained from the free energy density $f$ (our definition differs
from the usual one by a factor $\beta$), for instance
\eqn\feg{\rho=\partial_\beta f,\quad \sigma =\beta \rho -f.}
Obviously, to obtain formulas in \rst, it is enough to assume that
$f$ scales with $T$ as $\beta^2=T^{-2}$. For a noninteracting relativistic
gas, we have
\eqn\regas{f=\int_0^\infty F(\beta k)kdk,}
where $F$ as a function of $\beta k$ depends on the nature of the
constituents, bosons or fermions: $F=\mp\ln(1\pm \exp(-\beta k))$.
The measure $kdk$ in the momentum space
is standard for a 2 dimensional surface. In any case, \regas\ implies
that $f\sim T^2$, a familiar result. In order to have $f\sim T^{-2}$,
one can either replace the measure $kdk$ by $k^{-3}dk$, or to change
the dispersion relation. We have no reason to change the measure (a
consequence of quantum mechanics). Thus,
we need to change the dispersion relation $E=E(k)$ in $F(\beta E(k))$.
It is easy to see that if $E=m^2/k$, we get the desired result $f\sim
T^{-2}$. In conclusion, the following is forced upon us
\eqn\nfe{f=\int_0^\infty F(\beta m^2/k)kdk
=am^4\beta^2,}
with
\eqn\cons{a=\int_0^\infty F(k^{-1})kdk.}
The dimensionless constant $a$ is positive, unlike in the case of a
relativistic gas where it is negative. The mass $m$, according to
the discussion following \rst, is proportional to $m_P$.
The integral in \cons\ may be divergent if the behavior of the integrand
$F$ for large $k$ is similar to the Boltzman factor (as in the case of
a relativistic particle). To cure this problem, we need to postulate that
the phase space is regulated in such a way that for large $k$, the number
density of states becomes smaller, this helps to modify function $F$ to
make the integral convergent.

To summarize, the main result we have obtained so far is that the
basic constituent of the membrane fluid observes the dispersion
relation
\eqn\disp{E={m^2\over k},}
with $m\sim m_P$. This is a quite unusual relation, for $k=0$, the
energy of the quantum is infinity while for $k=\infty$ the energy
vanishes. Apparently, the usual Lorentz invariance breaks down on
the membrane, since the membrane is a special place where we probe
physics at Planck distances directly, there is no any contradiction.

One apparent interpretation of the above dispersion relation is
that on the membrane space and time are noncommuting. To see this,
remember that the usual quantum mechanics is still valid on membrane,
so momentum is still conjugate to space, and time is conjugate to
energy, thus $E\sim 1/\Delta T$ and $k\sim 1/\Delta X$, \disp\ implies
\eqn\stu{\Delta T\Delta X\sim m^{-2}\sim l_P^2.}
In the past, a spacetime uncertainty relation has been advocated \ty,
where the relevant scale is string scale. However, that relation is
relevant only when we are probing physics beyond string scale.
Relation \disp\ certainly suggests more radical revision of the spacetime
structure on the membrane, since on the face of it we do not
recover the usual dispersion relation in the low energy regime.

It is impossible to modify dispersion relation \disp\ in a way to
get the correct high energy behavior of the usual relativistic
dispersion relation. For instance, if we replace \disp\ by
$E=m^2/k +k$, then the contribution of large $k$ will make a correction
to \rst\ which is comparable to \rst. It is neither possible to
modify \disp\ to get the usual small $k$ dispersion relation
without spoiling \rst. For instance, if we replace \disp\ by
$E=m^2(k+m^2/k)^{-1}$, then the contribution of $k< m$ also
make a correction to \rst\ which can not be ignored for $T\sim
m_P$. Thus, we believe that the dispersion relation ought to be correct
for all range of $k$.

An alternative interpretation of integral \nfe\ is that the phase space
is not the two dimensional surface of the membrane (which we have omitted
thus far since we have been considering densities) together with their
canonical conjugate $k$. Rather, the phase space is the two dimensional
surface of the membrane and the size of a microscopic black hole.
These are the position of the fundamental quantum and its size, and,
the fact that there is no momentum of the quantum reflects the fact
that all constituents of the membrane and tightly packed. If we replace
$k$ by $1/L$, $L$ being the size of the black hole, then the dispersion
relation \disp\ is simply
\eqn\bhdisp{E=m^2L.}
For $m\sim m_P$, this is roughly the Schwarzschild relation between
the mass of the microscopic black and its size.

The original phase space measure $d^2xd^2k$, after factoring out $d^2x$
as in \nfe, becomes $kdk\sim L^{-3}dL$. A fundamental cell of the phase
space satisfies $\Delta^2 X\Delta L/L^3\sim 1$ or $\Delta^2 X\Delta L
\sim L^3$. The usual phase space cell is related to uncertainty relation,
here there are three variables, two coordinates and one size, so the
uncertainty relation is a triple one
\eqn\tunc{(\Delta X)^2\Delta L\ge L^3.}
We do not know how to explain this interesting uncertainty.

The interpretation of the fluid quantum as a microscopic black hole provides
a good reasoning for the proper distance $l$ being at the Planck scale.
We imagine that the membrane consists of microscopic black holes
tightly packed together, and for this to be the case, the size of a
typical hole is about $l$, otherwise the ``horizon" of the microscopic
hole will extend into the inside of the macroscopic hole. The size
of a typical hole is $L=E/m_P^2\sim T/m_P^2$, and $l=1/(2\pi T)$. If
$L\sim l$, we deduce $T\sim m_P$ and $l\sim l_P$.

Let us try to give further evidence for model of the microscopic black holes.
Let the typical size of a constituent black hole be $l$, and let it be a
free parameter, then the two dimensional
number density of the fluid is $n=c_1/l^2$ with a dimensionless parameter
$c_1$. Let the entropy carried by a microscopic black hole be $s$, which is
supposed of order $1$ as it should for a fundamental constituent, then the entropy
density is $\sigma=ns=(c_1s)/l^2$.
Since the mass of a typical microscopic black hole is of the order $l/l_p^2$,
thus the energy density of the fluid is $\rho\sim n l/l_p^2=c_1/(ll_p^2)$.
It follows that $\rho\sim \sqrt{\sigma}$ as functions of the free parameter
$l$, up to a constant depending on
$l_p$. We know that for a fluid with equation of state $p=w\rho$,
we have $\rho\sim \sigma^{1+w}$ (This relation can be derived using
two equations in \twor\ which are valid in general), we deduce that $w=-\half$, \
in exact agreement with \pres.

Of course, we may consider this strange fluid with a negative pressure extremely
suspicious, thus conclude that our discussion actually offers evidence
that the whole idea of the stretched horizon is wrong. Though this is a logic
possibility, we shall put aside this possibility and press ahead.

Although we have obtained the general form of the partition
function or free energy \nfe, we know nothing about the the
dynamic details of the membrane fluid. We may never know
much details without a fundamental theory of quantum gravity. Here
we shall try to see whether we can gain further information
about the fluid from the Hawking radiation \sh. The rate
for a black hole to emit a particle of spin $s$ and angular
momentum $l$ with energy $E$ is
\eqn\hra{{dN\over dt}=\int \sigma_{s,l} (E)(e^{\beta E}\mp 1)^{-1}
{E^2d E\over 2\pi^2},}
with $\sigma (E)$ is the greybody factor, $\mp$ sign depends
on whether the particle is a boson or a fermion, we also assumed
that the particle is massless.

Now, the number density of the fluid quantum is $g(\beta E)kdk$
with $g(x)=F'(x)$, $F$ is the function in \nfe. In terms of $E$,
this is simply $m^4g(\beta E)dE/E^3$. Since the black hole can
emit any kind of the particle existing in a fundamental theory,
we need to assume that the fluid quantum itself can transmute
into any kind of particle, regardless whether the particle is a boson or
a fermion. It is hard to imagine that any usual particle can be such
a quantum, only a microscopic black hole can play the role. If the
rate of the fluid quantum to transmute into a massless particle is
$A_{s,l}(E)$, then the radiation rate per unit area is
\eqn\rada{{dn\over dt}=m^4\int A_{s,l}(E)g(\beta E){dE\over E^3}.}
In the D-brane picture, the calculation of the decay rate on the
D-brane not only reproduces the Hawking radiation, it also reproduces
the greybody factor \ms, here we assume that the calculation performed
on the stretched membrane also reproduces the greybody factor. This is
to say that, the membrane not only encodes quantum information of the
black hole, it also knows about the global geometry of the black hole.
If so, then $A{dn\over dt}$ must be equal to formula \hra, we deduce
from this that
\eqn\ampl{A_{s,l}(E)={\sigma (E)\over A}{2\pi^2E^5\over m^4}
g^{-1}(\beta E)(e^{\beta E}\mp 1)^{-1}.}
Apparently, the factor $g^{-1}$ will never cancel the massless particle
thermal factor, so $A_{s,l}$ not only depends on the energy, it also
depends on the temperature. This result is not surprising, since
in order to compare \hra\ with \rada\ we need to blue-shift both $E$
and $T_H$ in \hra, and $T$ is at the Planck scale, thus the dependence
on $T$ of $A_{s,l}$ is simply a dependence on the Planck scale.
This dependence is natural for the dynamics of constituents involve
the Planck scale (for example the dispersion relation \disp).

To see that our idea is not one working accidentally in four dimensions,
in particular, to see the universality of the dispersion relation \disp,
we now switch to a Schwarzschild black hole in $d+2$ dimensions, here
$d$ is the dimensionality of the stretched membrane. The metric of such
a black hole is
\eqn\dsch{ds^2=-(1-{r_0^{d-1}\over r^{d-1}})dt^2+(1-{r_0^{d-1}\over r^{d-1}}
)^{-1}dr^2+r^2d\Omega^2_d.}
The relation between the horizon radius and the black hole mass is
\eqn\rerm{M={\Omega_d d\over 16\pi G}r_0^{d-1},}
where $G$ is the Newton constant in $d+2$ dimensions. The black hole
has a Hawking temperature
\eqn\dht{T_H={d-1\over 4\pi r_0}.}

Let the stretched membrane be a proper distance $l$ away from the real
horizon. The relation between the blue-shift factor and $l$ is
\eqn\bpre{g^{-1/2}_{00}(r_s)={2r_0\over (d-1)l},}
so the local temperature on the membrane is still
\eqn\lotem{T=T_Hg^{-1/2}_{00}(r_s)={1\over 2\pi l},}
and the total mass as seen by a local observer
\eqn\lomass{M(l)=Mg^{-1/2}_{00}(r_s)={d\over d-1}
{\Omega_d r_0^d\over 8\pi lG}.}
We shall see that the factor $d/(d-1)$ will play an important
role later.

On the membrane, the energy density is
\eqn\endensi{\rho={d\over d-1} {1\over 8\pi lG},}
and the entropy density
\eqn\entrd{\sigma ={1\over 4G}.}
The formula of the entropy density is universal, while
the one of the energy density is not. Relations in \twor\
are still valid, using them we find
\eqn\morth{p=-{1\over d}\rho,}
and
\eqn\moreth{\rho ={d\over d-1} cT^{-d+1},\quad \sigma=
cT^{-d}.}
In deriving the power $-d$ of $T$ in $\sigma$, the factor $d/(d-1)$
in \endensi\ played a role.

As in four dimensions, we postulate that the free energy is given
by
\eqn\freeen{f=\int F(\beta m^2/k)d^dk,}
and the right exponent of $T$ in $\sigma$ is obtained. Thus, the
dispersion relation in \disp\ is universal for stretched membranes
of the Schwarzschild black holes in all dimensions. Again, one can assume
the strange fluid quantum be just a microscopic black hole, and our
previous discussion applies to the general case, in particular,
all the microscopic holes are tightly packed, and their positions
and sizes satisfy a generalized uncertainty relation.

We believe that the local properties of the black hole membranes
are universal for Schwarzschild black holes, the Rindler space and even
the de Sitter space. In the last case, there is still disputation on
what is the total energy of the de Sitter space. The universality of
the stretched membrane suggests a formula. Let $l$ be the proper
distance between the stretched membrane and the real horizon of the
de Sitter space. The local temperature on the membrane is still
$T=1/(2\pi l)$. The relation between the red-shift factor (red-shift
between the membrane and the observer sitting at the origin) can be
found using the metric
\eqn\deme{ds^2=-(1-{r^2\over R^2})dt^2+(1-{r^2\over R^2})^{-1}dr^2
+r^2d\Omega^2_2,}
and is
\eqn\redsh{g^{1/2}_{00}(r_s)={R\over l}.}
If $\rho$ is the same as in \mmd, the total energy as seen by the
observer at $r=0$ is
\eqn\tote{E=\rho A g^{1/2}_{00}(r_s)={R\over l_p^2}.}
A similar formula can be obtained for a $d+2$ dimensional de Sitter
space using \endensi\ and \redsh.

Of course the suggestion that the properties of the black hole membranes
are universal may hold only for those cases when the heat capacity is
negative. An interesting example counters this suggestion is the AdS black
holes. For a large black hole whose size is larger than the AdS radius, the
heat capacity is positive, and it is well-known that the black hole is
described by the conformal field theory in which the dispersion relation
is the usual relativistic one. For small black holes, the heat capacity
becomes negative, their behavior is more like the Schwarzschild black holes.
Let us study the 5D AdS black holes in more details. The 5D AdS black hole
metric is
\eqn\adsbh{ds^2=-(1+{r^2\over R^2}-{r_0^2\over r^2})dt^2(1+{r^2\over R^2}-
{r_0^2\over r^2})^{-1}dr^2+r^2d\Omega^2_3.}
The physics of black holes described by the above metric has been studied
by many authors, for example, \gl\ studies the large N phase transition in
associated to the AdS black holes. For a given temperature, there are
black holes, one larger and one smaller, their radii are given by
\eqn\rdads{r_\pm ={\pi R^2T\over 2}\left(1\pm (1-{2\over (\pi RT)^2})^{1/2}
\right).}
And entropy
\eqn\endas{S_\pm ={\pi^2N^2\over 2}(2\pi^2R^3)T^3\left(\half \pm
\half (1-{2\over (\pi RT)^2})^{1/2}\right)^3,}
where we used the AdS/CFT correspondence to introduce the rank $N$ of the
gauge group in the conformal field theory.

We see from \endas\ that for the larger black hole, the leading
term in the entropy is proportional to $T^3$ and this is described
by a CFT gas, thus the relation between the energy density and the
entropy density is $\rho/(\sigma T)=3/4$ in the high temperature
thus very large black hole limit, since we have a 3D gas (for a 4D
AdS black hole, the ratio becomes $2/3$). However, for the smaller
black hole in the high temperature limit, the leading term in
entropy is proportional to $T^{-3}$, this is the behavior of a 5D
Schwarzschild black hole. It is interesting to study further how
the CFT physics describes the transition between the larger black
hole and the smaller black hole, thus gaining insight into physics
of the Schwarzschild black holes.

To conclude, starting from the standard black hole thermodynamics and
the postulate of the stretched membrane, we find that spacetime are
noncommutative on the membrane. It is possible to further explore
quantum properties of the stretched membrane based on this result.

Acknowledgments.
This work was supported by a
``Hundred People Project'' grant of Academia Sinica and an outstanding
young investigator award of NSF of China. This work was inspired by discussions
during a tea party in Hefei, the Interdisciplinary Centers of Theoretical
Sciences of USTC and of CAS are gratefully acknowledged. I also thank many friends who
attended that tea party, in particular R. G. Cai and J. X. Lu for discussions.
F. L. Lin is thanked for his comment on the manuscript.

\listrefs
\end